\documentclass[aps,pre,reprint,floatfix]{revtex4-2}

\usepackage{amsmath}
\usepackage{graphicx}
\usepackage[separate-uncertainty]{siunitx}
\usepackage{xcolor}

\graphicspath{{figures/}}

\newcommand{\comment}[1]{}

\begin{document}

\title{Acoustodynamic mass determination:
Accounting for inertial effects in acoustic levitation of granular materials}

\author{Mia C. Morrell}
\author{David G. Grier}

\affiliation{Department of Physics and Center for Soft Matter Research, New York University, New York, NY 10003, USA}

\begin{abstract}
    Acoustic traps use forces exerted by sound waves to confine and transport small objects.
    The dynamics of an object moving in the force landscape of an acoustic trap can be significantly influenced by the inertia of the surrounding fluid medium.
    These inertial effects can be observed by setting a trapped object in oscillation and tracking it as it relaxes back to mechanical equilibrium in its trap.
    Large deviations from Stokesian dynamics during this process can be explained quantitatively by accounting for boundary-layer effects in the fluid.
    The measured oscillations of a perturbed
    particle then can be used not
    only to calibrate
    the trap but also to characterize the
    particle.
\end{abstract}

\maketitle

\section{Introduction}

Acoustic manipulation of granular media
was first demonstrated by Kundt in 1866 as a
means to visualize the nodes and antinodes of sound waves \cite{kundt1866about}.
After a century and a half of gestation,
acoustic trapping is emerging as a
focal area for soft-matter physics
\cite{lim2019cluster,kline2020precision,mendez2022lifetime,lim2023acoustically}
and a practical
platform for dexterous noncontact materials
processing \cite{foresti2013acoustophoretic,andrade2020acoustic}
thanks in part to recent advances
in the theory of wave-matter interactions
\cite{bruus2012acoustofluidics2,abdelaziz2020acoustokinetics} and
innovations in the techniques for crafting
acoustic force landscapes
\cite{marzo2017tinylev,marzo2019holographic}.
An object's trajectory through such a
landscape encodes information
about the wave-matter interaction and therefore can
be used not just to calibrate
the trap but also to characterize the object.
The present study demonstrates how to extract
that information
through machine-vision measurements
of trapped objects' oscillations under the combined influences
of gravity, the trap's restoring
force and drag due to displacement of the
surrounding fluid medium.

Correctly interpreting
the measured trajectory of an acoustically
trapped particle can be
challenging because
the drag force deviates substantially from
the standard Stokes form,
as has been noted in previous studies
\cite{perez2014experimental,andrade2014experimental,nakahara2022acoustic,marrara2023optical}.
We incorporate
non-Stokesian drag into a self-consistent measurement framework by invoking
Landau's hydrodynamic boundary-layer approximation \cite{landau1987fluids,settnes2012forces} to account for the
fluid's inertia.
This approach appears not to have been demonstrated previously and provides
a fast and accurate way to measure physical
properties
of the trapped object without requiring separate
calibration of the acoustic trap.
The same measurement also yields an absolute
calibration of the trap's stiffness for that
specific object.

\begin{figure}
    \centering
    \includegraphics[width=\columnwidth]{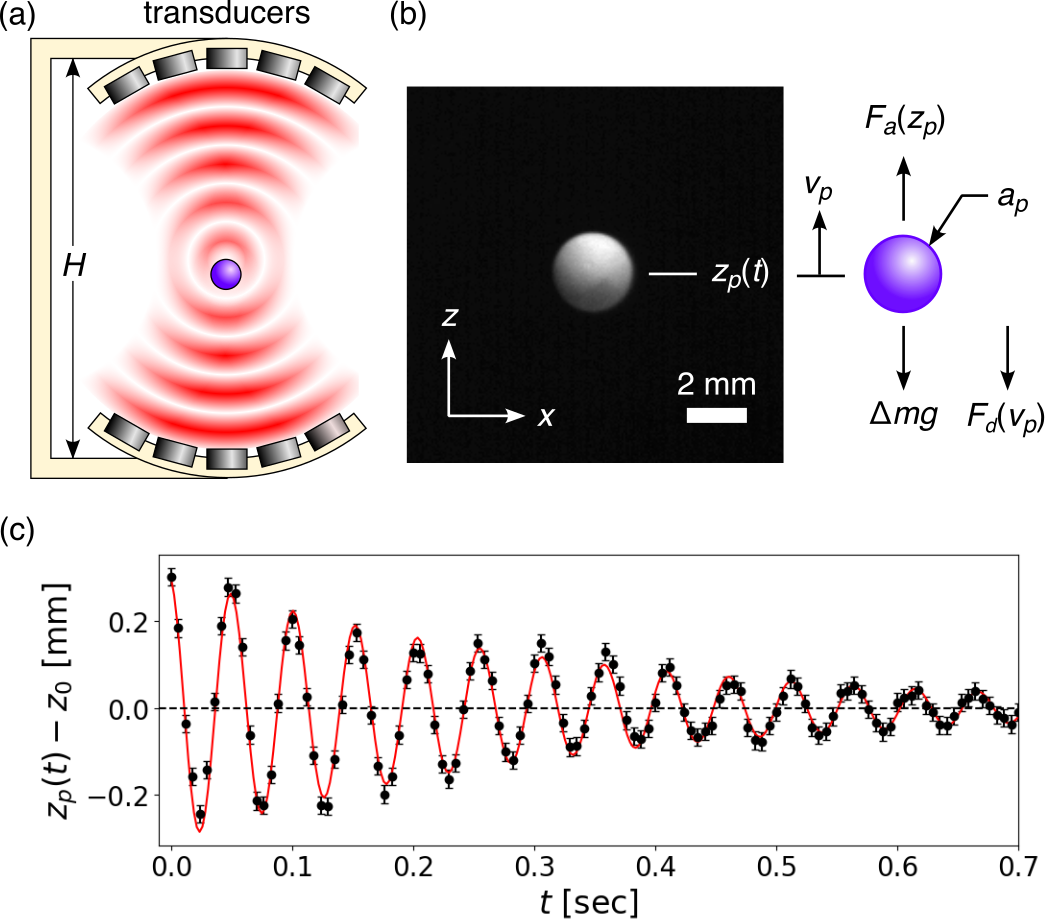}
    \caption{(a) Schematic diagram of the reference acoustic trap. (b) Typical video frame of a millimeter-scale styrofoam sphere
    levitated in the acoustic trap together
    with a schematic diagram of the
    forces acting on the particle.
    (c) Measured trajectory (black symbols)
    of a styrofoam bead returning
    to mechanical equilibrium in an acoustic trap compared with predictions of the damped
    oscillator model (red curve).}
    \label{fig:oscillate}
\end{figure}

\section{Dynamics of an Acoustically Trapped Particle}

\subsection{Imaging measurements of damped oscillations}

Figure~\ref{fig:oscillate}(a) schematically represents
the acoustic trapping system used for this study.
Based on the standard TinyLev design \cite{marzo2017tinylev}, this acoustic levitator
consists of two
banks of piezoelectric ultrasonic transducers
(MA40S4S, Murata, Inc.)
with a resonance frequency around \SI{40}{\kilo\hertz}.
Each bank of \num{36} transducers
is driven sinusoidally
by a function generator
(DS345, Stanford Research Systems)
and projects a traveling wave
into a spherical volume of air.
Interference between the two waves creates
an array of acoustic traps along the instrument's vertical axis.
Figure~\ref{fig:oscillate}(b) presents a video
image of a millimeter-scale sphere
of expanded polystyrene localized in air
within one of the acoustic traps.
The camera (Blackfly S USB3, FLIR) records
the particle's motions at \SI{170}{frames\per\second}.
with an exposure time of \SI{2}{\ms} and
an effective magnification of
\SI{61}{\um\per pixel}.
Under these imaging conditions, the height
of the particle in the trap, $z_p(t)$,
can be measured in the
imaging plane to within
$\epsilon_z = \SI{10}{\um}$ by fitting
for the image's least bounding circle
\cite{gonzalez2008digital}.
This method has the advantage over
light-scattering techniques
\cite{marrara2023optical} that it
also yields an estimate for the
particle's radius, $a_p$.
For the particle in Fig.~\ref{fig:oscillate}(b),
$a_p = \SI{1.346(7)}{\mm}$.

The particle can be made to oscillate in its
trap by rapidly displacing it from its
equilibrium position. This can be accomplished
by abruptly changing the amplitude, frequency
or relative phase \cite{nakahara2022acoustic}
of the signals driving the two banks of
transducers.
The discrete symbols in
Fig.~\ref{fig:oscillate}(c) show the
trapped particle's displacement from its
equilibrium position, $z_p(t) - z_0$,
after an abrupt change of drive amplitude
causes a displacement of
$\Delta z = \SI{-0.31(1)}{\mm}$.
The (red) curve is a fit to the standard
result for a damped harmonic oscillator,
\begin{equation}
    \label{eq:fit}
    z_p(t)
    =
    z_0
    +
    \Delta z \, e^{-\frac{1}{2} \gamma t} \cos(\Omega t)
\end{equation}
for the oscillation frequency,
$\Omega = \SI{122.1(2)}{\radian\per\second} =
\SI{19.43(3)}{\hertz}$
and the damping rate
$\gamma = \SI{6.3(3)}{\per\second}$,
in addition to $\Delta z$ and the
equilibrium position, $z_0$.
Although this fit appears to be satisfactory,
the estimated parameters must be interpreted
with care.

To illustrate the challenge, consider
the standard Stokes result,
\begin{equation}
\label{eq:stokesdrag}
    \gamma_0 = \frac{6 \pi \eta_m \, a_p}{m_0},
\end{equation}
for the drag rate experienced by a sphere
of radius $a_p$ and mass $m_0$
as it moves through a fluid
with dynamic viscosity $\eta_m$.
Expanded polystyrene sphere has a
density of roughly
$\rho_p = \SI{30}{\kilogram\per\cubic\meter}$
\cite{horvath1994expanded},
so that
\begin{equation}
    m_0 = \frac{4}{3} \pi a_p^3 \, \rho_p
    = \SI{0.3}{\mg}.
\end{equation}
The viscosity of air under standard
conditions is $\eta_m = \SI{1.825(5)e-5}{\pascal\second}$ \cite{rumble2023crc}.
Equation~\eqref{eq:stokesdrag} therefore
predicts
$\gamma_0 = \SI{1.4}{\per\second}$, which
is a factor of four smaller than the measured
value.
Previous studies on similar systems have
reported comparably large discrepancies between
predicted and observed drag rates
\cite{marrara2023optical}, and
have addressed them phenomenologically
with nonlinear drag models
\cite{perez2014experimental,fushimi2018nonlinear,marrara2023optical}, if at all
\cite{andrade2014experimental,nakahara2022acoustic}.

Here, we demonstrate that
the linear drag model underlying Eq.~\eqref{eq:fit}
indeed is appropriate for analyzing
the oscillations of acoustically
trapped objects provided that
the parameters are suitably modified
to account for
the inertia of the displaced fluid
\cite{landau1987fluids,settnes2012forces}.
The enhanced model provides a basis for
precisely measuring the
density and mass of trapped objects
without requiring the acoustic trap
to be independently calibrated.
The same approach can be used to calibrate the
trap's stiffness while accounting naturally
for the influence of external forces such
as gravity.

\subsection{Acoustic forces}

The force landscape experienced by an
acoustically trapped object is dictated by the
structure of the sound field.
The counterpropagating
waves in our instrument
interfere to create a standing
pressure wave along the central axis
whose spatial dependence
is approximately sinusoidal,
\begin{equation}
    p(z)
    =
    2 p_0 \sin(k z),
\end{equation}
near the midplane at $z = 0$.
Here, $p_0$ is the pressure amplitude
due to a single bank of transducers and
$k = \omega/c_m$ is the wave number of
sound at frequency $\omega$
in a medium whose speed of sound is $c_m$.
For acoustic levitation in air,
$c_m = \SI{343.5(5)}{\meter\per\second}$
under standard conditions \cite{rumble2023crc}.

The time-averaged acoustic force experienced
by an object
at height $z_p$ in the standing
wave has the form \cite{bruus2012acoustofluidics2,abdelaziz2020acoustokinetics}
\begin{equation}
    F_a(z_p)
    =
    F_0 \sin(2 k z_p),
\end{equation}
with an overall scale,
\begin{equation}
\label{eq:forcescale}
    F_0 = \chi \, k p_0^2,
\end{equation}
that depends on the frequency and amplitude of
the sound wave, and on properties of the object
and the medium through $\chi$.
For a small spherical particle ($k a_p < 1$),
the acoustic response function is
\cite{bruus2012acoustofluidics2,abdelaziz2020acoustokinetics}
\begin{equation}
    \chi =
    \frac{4}{3} \pi a_p^3 \,
    \kappa_m
    \left(
    1 - \frac{\kappa_p}{\kappa_m}
    +
    3 \frac{\rho_p - \rho_m}{2 \rho_p + \rho_m}
    \right),
\end{equation}
where $\rho_p$ and $\rho_m$ are the densities
of the particle and medium, respectively,
and $\kappa_p$ and $\kappa_m$ are their
respective isentropic compressibilities.
Dense incompressible particles
have $\chi > 0$
and therefore tend to be
trapped at nodes of the pressure field.

External forces can
displace the particle from the
center of the acoustic trap,
as depicted in Fig.~\ref{fig:oscillate}(b).
Gravity, in particular, acts on the particle's
buoyant mass,
\begin{equation}
    \Delta m
    =
    m_0 \left(1  - \frac{\rho_m}{\rho_p}\right).
\end{equation}
and displaces it from the
pressure node at
$z = 0$ into
mechanical equilibrium at
\begin{equation}
\label{eq:z0}
    z_0
    =
    - \frac{1}{2 k}
    \sin^{-1} \left(
    \frac{\Delta m g}{F_0} \right),
\end{equation}
where $g = \SI{9.81}{\meter\per\square\second}$
is the acceleration due to gravity.

As long as the particle does not move
too far from the nodal plane,
the acoustic trap
exerts an approximately Hookean
restoring force on the particle,
\begin{equation}
    F_a(z_p)
    \approx
    - \kappa
    \, (z_p - z_0),
\end{equation}
with a stiffness,
\begin{equation}
\label{eq:kappa}
    \kappa
    =
    2 k F_0 \, \cos(2 k z_0),
\end{equation}
that depends on properties of the
sound wave, properties of the particle
and the strength of the external force.
Calibrating the trap generally involves
determining $\kappa$.
Equation~\eqref{eq:kappa} clarifies that
the trap cannot be calibrated
with a reference object, as has been proposed \cite{nakahara2022acoustic},
but instead requires a separate calibration
for every set of experimental
conditions.

\subsection{Inertial corrections}

An object moving in the trap's potential energy
well displaces the surrounding fluid medium and
therefore experiences viscous drag.
The standard Stokes result from Eq.~\eqref{eq:stokesdrag}
neglects the inertia of the fluid.
For the special case of a sphere undergoing
harmonic oscillations,
inertial effects
can be incorporated into a linear drag model,
\begin{equation}
    F_d(v_p)
    =
    - m(\Omega) \,  \gamma(\Omega) \, v_p,
\end{equation}
by defining a dynamical mass
\begin{equation}
\label{eq:dynamicmass}
    m(\Omega)
    =
    m_0 \left(
    1 + \frac{1}{2} \frac{\rho_m}{\rho_p}
    \left[ 1
    + \frac{9}{2} \frac{\delta(\Omega)}{a_p} \right]
    \right)
\end{equation}
and a renormalized drag rate,
\begin{equation}
\label{eq:dragrate}
    \gamma(\Omega)
    =
    \frac{6 \pi \eta_m a_p}{m(\Omega)}
    \left(1 + \frac{a_p}{\delta(\Omega)}\right)
\end{equation}
that both depend on the oscillation
frequency, $\Omega$, through the
thickness of a boundary layer surrounding the sphere \cite{landau1987fluids,settnes2012forces},
\begin{equation}
\label{eq:delta}
    \delta(\Omega)
    =
    \sqrt{\frac{2 \eta_m}{\rho_m}
    \frac{1}{\Omega}} .
\end{equation}
The inertia-corrected equation of motion for
an acoustically levitated sphere
is then analogous to the standard
equation of motion for
the damped harmonic oscillator,
\begin{equation}
\label{eq:equationofmotion}
    \ddot{z}_p +
    \gamma(\Omega) \, \dot{z}_p +
    \Omega_0^2 \, (z_p - z_0) = 0,
\end{equation}
with a natural frequency,
\begin{equation}
\label{eq:omega0}
    \Omega_0(\Omega) = \sqrt{\frac{\kappa}{m(\Omega)}},
\end{equation}
that is related to the measured frequency by
\begin{equation}
    \label{eq:frequencyrelation}
    \Omega_0^2(\Omega)
    =
    \Omega^2
    +
    \frac{1}{4} \gamma^2(\Omega).
\end{equation}
Unlike the standard harmonic oscillator,
whose drag and restoring forces are independent
of frequency, the natural frequency
of an acoustically trapped object
must be found by solving
Eq.~\eqref{eq:frequencyrelation}
self-consistently.

The derivation
of Eq.~\eqref{eq:equationofmotion}
from boundary-layer theory
establishes that Eq.~\eqref{eq:fit}
suitably models
the dynamics of an
object oscillating in an acoustic trap.
Unlike dynamical models with nonlinear drag
\cite{perez2014experimental,marrara2023optical},
the damping rate predicted by Eq.~\eqref{eq:dragrate}
does not depend on
the amplitude of the motion.
This is consistent with the observation
in Fig.~\ref{fig:oscillate}(c) that a
single constant value for $\gamma$
successfully accounts for viscous damping
over the oscillating particle's entire trajectory.

\begin{figure*}
    \centering
    \includegraphics[width=0.9\textwidth]{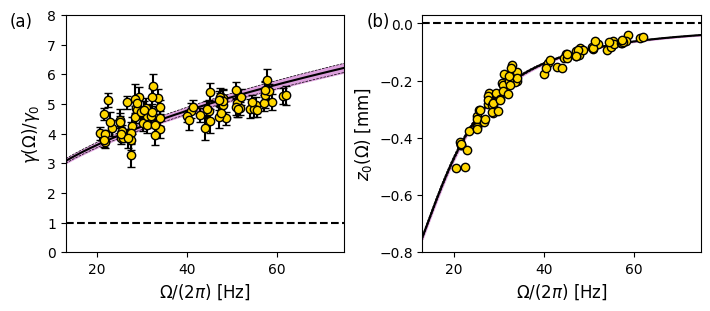}
    \caption{(a)
    Frequency dependence of the
    damping rate $\gamma(\Omega)$,
    Plot symbols present results
    from fits to measured trajectories, such as the
    example in Fig.~\ref{fig:oscillate}(c).
    The solid curve is a fit to the boundary-layer
    model in Eq.~\eqref{eq:dragrate}
    that yields
    $\rho_p = \SI{28.9(3)}{\kg\per\cubic\meter}$.
    The horizontal dashed line reflects the
    standard Stokes result,
    $\gamma_0 = \SI{1.50(3)}{\per\second}$.
    (b) Correlation of the axial offset, $z_0$,
    with the observed
    oscillation frequency, $\Omega$.
    The horizontal dashed line represents the
    nodal plane of the acoustic trap.}
    \label{fig:gamma}
\end{figure*}

\section{Acoustodynamic Mass Determination}

Measurements of $\gamma(\Omega)$
can be interpreted with Eq.~\eqref{eq:dragrate}
to estimate the mass density of the particle.
The discrete points in Fig.~\ref{fig:gamma}(a)
are measured by fitting recorded trajectories
of the expanded
polystyrene bead in Fig.~\ref{fig:oscillate}.
Different oscillation frequencies are obtained by
adjusting the the amplitude, $V_0$, of the sinusoidal voltage powering the trap.
It is not necessary to know how the
trap strength, $F_0$, depends on $V_0$
to perform this measurement because $\gamma$
and $\Omega$ are both obtained directly from
each measured trajectory.
Taking the density of air to be
$\rho_m = \SI{1.220(5)}{\kg\per\cubic\meter}$
\cite{rumble2023crc} leaves the
particle's density,
$\rho_p$, as the only undetermined parameter in the model.
The solid curve in Fig.~\ref{fig:gamma}(a)
is a fit to Eq.~\eqref{eq:dragrate} that
yields $\rho_p = \SI{28.9(3)}{\kg\per\cubic\meter}$,
which is consistent with expectations for
expanded polystyrene beads \cite{horvath1994expanded}.
Inertial corrections quite convincingly
account for the previously unexplained enhancement
of the oscillating particle's drag rate.
In so doing, they also
provide the basis for a precise and robust
way to measure the mass density of millimeter-scale objects.
Combining $\rho_p$ with the optically-measured radius
yields the levitated object's mass,
$m_0 = \SI{0.31(1)}{\mg}$.
Repeating this measurement on
\num{10} different beads from the same batch
yields an average density of $\rho_p = \SI{30.5(2)}{\kg\per\cubic\meter}$ and an average mass
of $m_0 = \SI{0.295(3)}{\mg}$.

The precision of acoustodynamic
mass determination is
limited by run-to-run variability in
the measured values of $\gamma(\Omega)$,
which in turn can be ascribed
to spurious transverse motions of the particle
in its trap
and to environmental factors such as vibrations and drafts.
Even with these practical limitations,
the \SI{3}{\ug} precision achieved in
this representative realization is
comparable to the performance of a conventional
ultra-micro balance.

Previous acoustic trapping
studies have attempted to measure
the masses of levitated objects
by interpreting their static displacements
\cite{trinh1986acoustic} with Eq.~\eqref{eq:z0}
or by interpreting their oscillation frequencies
directly \cite{nakahara2022acoustic} without inertial corrections.
Like conventional scales and balances,
these approaches rely on independent calibration
of the trap's stiffness, $\kappa$.
The present acoustodynamic approach
avoids the need for such calibrations by
comparing two independent time scales represented
by $\Omega$ and $\gamma$,
rather than two independent force scales.

Increasing the acoustic trap's strength increases
the oscillation frequency and lifts the
particle toward the trap's center.
This correlation is reflected in the
dependence of $z_0$ on $\Omega$ that is plotted in Fig.~\ref{fig:gamma}(b).
These measurements
can be interpreted within the boundary-layer
model by
combining Eq.~\eqref{eq:z0} with Eq.~\eqref{eq:kappa}
to obtain
\begin{equation}
    \label{eq:z0prediction}
    z_0(\Omega)
    =
    -\frac{1}{2k} \,
    {\tan^{-1}}{\left(
    \frac{2k g}{\Omega_0^2(\Omega)}
    \frac{\Delta m}{m(\Omega)}
    \right)} + z_\text{trap},
\end{equation}
where $z_\text{trap}$ is the height of the
trap's nodal plane in the camera's field of view.
The solid curve in Fig.~\ref{fig:gamma}(b)
shows this model's prediction using
the value of $\rho_p$ obtained from $\gamma(\Omega)$.
The data in Fig.~\ref{fig:gamma}(b)
have been offset so
that $z_\text{trap} = \SI{0(1)}{\um}$.
The excellent agreement between measurement and
theory in this comparison serves to validate
the acoustodynamically determined
values of $\rho_p$ and $a_p$.
Accurately identifying $z_\text{trap}$ also is
valuable for force-extension measurements
once the trap's stiffness is calibrated.

\section{Dynamic Trap Calibration}

\begin{figure}
    \centering
    \includegraphics[width=0.9\columnwidth]{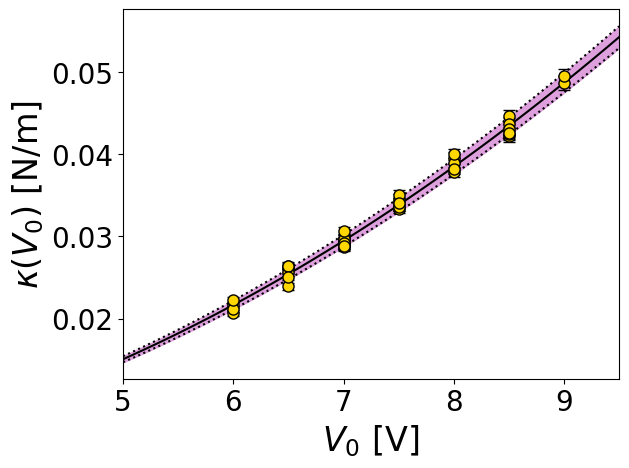}
    \caption{Dependence of the measured trap
    stiffness, $\kappa$, on the peak-to-peak
    voltage, $V_0$, used to power the acoustic
    trap's transducer banks.
    The solid curve is a fit to Eq.~\eqref{eq:calibration} for the
    calibration constant, $\alpha$.
    }
    \label{fig:calibration}
\end{figure}

The trap's stiffness at each value of $V_0$
can be inferred from the
particle's damped oscillations through
\begin{subequations}
    \label{eq:trapcalibration}
\begin{equation}
\label{eq:kappav}
    \kappa(\Omega) =
    m(\Omega) \, \Omega_0^2(\Omega).
\end{equation}
Assuming that
the pressure amplitude, $p_0$,
is proportional
to the driving voltage, $V_0$,
Eq.~\eqref{eq:forcescale}
and Eq.~\eqref{eq:kappa}
lead to an independent expression,
\begin{equation}
\label{eq:calibration}
    \kappa(V_0)
    \approx
    \alpha V_0^2 \
    \left[
    1 -
    \frac{1}{2}
    \left(\frac{k \Delta m g}{\alpha V_0^2}\right)^2
    \right] ,
\end{equation}
\end{subequations}
that can be compared with
measurements based on
Eq~\eqref{eq:kappav}
to obtain $\alpha$, the required
calibration constant
for this specific particle in the
levitator.
This result is valid when the particle is
stably trapped against gravity, $F_0 > \Delta m g$.
Figure~\ref{fig:calibration} shows the
calibration obtained from the data set
in Fig.~\ref{fig:gamma} and yields
$\alpha = \SI{6.02(15)e-4}{\newton\per\meter\per\square\volt}$.
Ignoring inertial corrections
by using $m_0$ in Eq.~\eqref{eq:kappav}
would have yielded a significant
underestimate for the calibration constant,
$\alpha_0 = \SI{5.77(14)e-4}{\newton\per\meter\per\square\volt}$.
Adding to the challenge, $m_0$ generally
would not be known
accurately for a millimeter-scale object
\emph{a priori}.
The analytical framework described here solves this
problem by providing self-consistent measurements
of $\rho_p$, $m_0$ and $m(\Omega)$.
The resulting calibration constant, $\alpha$, therefore should yield
reliable results for the
trap stiffness, $\kappa$.

\section{Discussion}

Abruptly changing the trapping characteristics
of an acoustic levitator sets a trapped
object into a free oscillation that is
damped by viscous drag in the surrounding
medium.
The resulting trajectory can be
described with the standard model for a damped
harmonic oscillator provided that
inertial effects in the displaced fluid
are taken into account self-consistently
with hydrodynamic boundary-layer theory
\cite{landau1987fluids,settnes2012forces}.
These inertial corrections quantitatively
resolve the large discrepancy between
the measured drag rate and the Stokes prediction
that has been noted in previous studies, but previously
has been unexplained.
The boundary-layer model is applicable
for particle speeds substantially smaller than the
speed of sound and Reynolds numbers well
below the threshold for turbulence.
For the present study, $v_p < \SI{0.25}{\meter\per\second}$ and
$\text{Re} = \rho_p a_p v_p / \eta_m \lesssim \num{23}$, so that
both conditions are satisfied.
The observation that $\text{Re} > 1$ explains
why the standard Stokes result substantially underestimates
the drag rate.

Fitting measured trajectories to predictions
of the boundary-layer model yields precise
estimates for the trapped particle's mass density
and mass. Dynamic acoustic trapping
therefore can be used to weigh
millimeter-scale objects without requiring
direct contact, including sub-milligram
objects that can be challenging to weigh individually
\cite{shaw2016milligram,shaw2018current,ota2021evaluation}.
Generalizing this approach to accommodate
aspherical objects, powders and fluids
will be addressed in future studies.

Using the acoustic force field itself to set an object
into oscillation provides a simple and
effective method to calibrate the
stiffness of an acoustic trap.
This approach does not require the external
intervention used in complementary
calibration techniques, such as mechanically
moving the sample relative to the levitator
\cite{li2013simple,lim2018evaluation}.
The techniques discussed in this work therefore
should facilitate fundamental research on the
dynamics of granular materials in
acoustic force landscapes.

Acoustodynamic mass determination
should have near-term applications in the
pharmaceutical industry for weighing individual
pills and capsules, in the jewelry industry for weighing gemstones
and precious metals, and in the nuclear power
industry for massing
individual fuel pellets.
Many such applications currently rely on ultra-micro balances
to cover the relevant mass range with good precision.
Acoustodynamic mass determination offers several advantages.
The measurement is inherently self-calibrated and
is robust against environmental perturbations.
Levitated samples never come in contact with surfaces,
which is inherently beneficial for sensitive and
hazardous materials, minimizes
the likelihood of cross-contamination
and simplifies integration with robotic sample handlers.
Unlike conventional techniques,
furthermore, acoustodynamic mass
determination can operate freely in
challenging environments such as microgravity.

\section*{Acknowledgements}

This work was supported by the National Science
Foundation under Award No.~DMR-2104837.
The authors acknowledge helpful conversations
with Marc Gershow.

% \bibliography{calibrate}

\end{document}